\documentclass[12pt,a4paper]{article}
\pdfoutput=1

\usepackage[utf8]{inputenc}
\usepackage{amsmath}
\usepackage{amsfonts}
\usepackage{amssymb}
\usepackage{algorithmicx}
\usepackage{algorithm2e}
\usepackage{cite}
\usepackage{amsthm}
\usepackage{breqn}
\usepackage{titling}
\usepackage{lipsum}
\usepackage{authblk}

\newcommand{\N}{\mathbb{N}}

\begin{document}

\title{A Formula That Generates Hash Collisions}
\author{Andrew Brockmann}
\affil{University of Wisconsin, Madison \\ Computer Sciences Department \\ abrockmann@wisc.edu}
%\date{} % Uncommenting this line removes the date

\maketitle

\begin{abstract}
We present an explicit formula that produces hash collisions for the Merkle-Damg\aa rd construction. The formula works for arbitrary choice of message block and irrespective of the standardized constants used in hash functions, although some padding schemes may cause the formula to fail. This formula bears no obvious practical implications because at least one of any pair of colliding messages will have length double exponential in the security parameter. However, due to ambiguity in existing definitions of collision resistance, this formula arguably breaks the collision resistance of some hash functions.
\end{abstract}

\section{Introduction}

Collision resistance is one of the fundamental desirable properties of a hash function. Since hash values are often used as digital fingerprints, it should be difficult to find pairs of messages with the same hash value.

The main contribution of this paper is a formula that deterministically produces partial or full collisions for Merkle-Damg\aa rd hash functions, such as MD5, SHA1, and the SHA2 family. The formula provides collisions on the iterated compression function for any Merkle-Damg\aa rd hash function. It may or may not provide full collisions depending on the padding scheme and limitations on input length. In the case of MD5, the formula yields full collisions. It does not provide full collisions for SHA1 and SHA256 due to the requirement that inputs have fewer than $2^{64}$ bits.

After explaining the formula, we show that it produces messages with length double exponential in the security parameter, and that this methodology probably cannot be used to find colliding messages of sub-exponential length. However, the colliding messages are highly redundant and can be compressed to logarithmic space. Because of ambiguity in the existing definitions of collision resistance, we point out that one can arguably perform a trivial collision attack by simply writing out the collision formula.

\section{Background}

The Merkle-Damg\aa rd construction has been used to design some of the most commonly used hash functions. The construction is conceptually simple: the input message is read one block at a time to iteratively update an internal state, and after all blocks have been read, the final internal state is used to derive the hash value.

More specifically, the construction specifies a block length $b$, an output length $n$, an internal state size $\ell$, an initial internal state $IV$, and a compression function $f: \{0,1\}^b \times \{0,1\}^\ell \to \{0,1\}^\ell$. The following steps are performed to determine the hash value:
\begin{itemize}
\item[1.] Pad the input so that the total length is a multiple of the block length.
\item[2.] Initialize the internal state $IV$.
\item[3.] Read in the individual blocks of the input; for each block $B$, update $IV$ as
	\[IV \leftarrow f(B, IV).\]
\item[4.] Once all blocks have been read, derive the hash value from the final value of $IV$.
\end{itemize}
Note that the inner workings of the construction are much like a DFA, with states corresponding to the possible values of $IV$ and transitions determined by the blocks of the padded message.

The functions MD5, SHA1, and SHA256 share some similarities: all three use $b = 512$, let $\ell = n$, and simply output the final value of $IV$ as the hash value. They also use the same padding scheme (up to endianness), which works as follows:
\begin{itemize}
\item[1.] Append a single ``1'' bit to the input.
\item[2.] Append ``0'' bits until the total length is equivalent to 448 modulo 512.
\item[3.] Append the input length as a 64-bit integer, producing a padded message with length a multiple of 512.
\end{itemize}
If the input has $2^{64}$ or more bits, then its length cannot be encoded in a 64-bit integer. In this case, MD5 simply uses the lowest order 64 bits of the input length \cite{md5}. The NIST standardizations of SHA1 and SHA256, meanwhile, prohibit these long inputs \cite{fips180}.

If we fix the block $B \in \{0,1\}^b$ in the compression function $f$, the result is a function $f_B: \{0,1\}^\ell \to \{0,1\}^\ell$. An ideal hash function should be difficult to distinguish from a random function; therefore, if a hash function is well designed, then we expect $f_B$ to behave as a randomly chosen function from $\{0,1\}^\ell$ to itself. Consequently, hash functions are sometimes studied by examining the expected asymptotic properties of random mappings on finite domains. We now define and state some of these properties. \\

\noindent \textbf{Definition 1:} Suppose $X$ is a finite set, and let $f: X \to X$. The \emph{functional graph} of $f$ is the directed graph $G_f$ with vertex set
	\[V = X\]
and edge set
	\[E = \{(x, f(x)) ~|~ x \in X\}.\]
	
In general, each component of any functional graph will have a central directed cycle and some directed trees leading to the cycle \cite{HAC}. Next, we define some of the properties of nodes in a functional graph: \\

\noindent \textbf{Definition 2:}  Suppose $f: X \to X$. For any $x \in X$:
\begin{itemize}
\item[1.] The \emph{tail length} $\lambda(x)$ of $x$ is the minimum length of a directed path from $x$ to a cyclic node in $G_f$.
\item[2.] The \emph{cycle length} $\mu(x)$ of $x$ is the length of the cycle in the component containing $x$.
\item[3.] The \emph{rho length} $\rho(x)$ of $x$ is the sum of its tail and cycle lengths, $\rho(x) = \lambda(x) + \mu(x)$.
\end{itemize}
Following are some expected properties of individual nodes in random functional graphs, as given in the \emph{Handbook of Applied Cryptography}: \\

\noindent \textbf{Theorem 1 \cite{HAC}:} Suppose $|X| = N$. If $x \in X$ and $f: X \to X$ are chosen uniformly at random, then the expected tail, cycle, and rho lengths of $x$ are asymptotically given as follows:
\begin{itemize}
\item[1.] $\lambda(x) \sim \sqrt{\pi N/8}$
\item[2.] $\mu(x) \sim \sqrt{\pi N/8}$
\item[3.] $\rho(x) \sim \sqrt{\pi N/2}$
\end{itemize}
Finally, we state expectations for some properties of the functional graph itself: \\

\noindent \textbf{Theorem 2 \cite{HAC}:} Suppose $|X| = N$, and $f: X \to X$ is chosen uniformly at random. Then the following are asymptotically true as $N \to \infty$:
\begin{itemize}
\item[1.] The expected number of cyclic nodes in $G_f$ is $\sim\sqrt{\pi N/2}$.
\item[2.] The expected maximum tail length of a node in $G_f$ is $\sim c_1\sqrt{N}$, where $c_1 \approx 0.78$.
\item[3.] The expected maximum cycle length in $G_f$ is $\sim c_2\sqrt{N}$, where $c_2 \approx 1.74$.
\item[4.] The expected maximum rho length in $G_f$ is $\sim c_3\sqrt{N}$, where $c_3 \approx 2.41$.
\end{itemize}
While nodes in functional graphs can have variable in-degree, each has out-degree 1, since $f$ is a function. Starting from any node in any $G_f$, traversing the out-edges will eventually lead us to a cycle.

Suppose now that we let $X = \{0,1\}^\ell$, the set of possible internal states for a hash function, and choose $f$ to be the the function $f_B$ obtained by fixing a block $B$ in the hash compression function. Let $[B]^k$ denote the message consisting of the block $B$ repeated $k$ times. Observe that if we begin at the node $IV$ in the functional graph $G_{f_B}$ and traverse $k$ out-edges, we arrive at the node
	\[IV' = f_B^k(IV) = f_B(f_B(\dots f_B(IV))).\]
The value $IV'$ is also the intermediate internal state of a hash function after processing $k$ blocks that are all identical to $B$. This observation, and the aforementioned facts about functional graphs, inspire our collision formula.

\section{The Collision Formula}

Arbitrarily fix a block $B \in \{0,1\}^b$ in a hash function's compression function to obtain a function $f_B: \{0,1\}^\ell \to \{0,1\}^\ell$. Consider the messages
	\[M_0 = [B]^{\lambda(IV)}, ~~~~~ M_1 = [B]^{\lambda(IV) + \mu(IV)}.\]
When $f_B$ is applied iteratively to these messages one block at a time, the two clearly agree after the first $\lambda(IV)$ blocks, since the messages are identical up to that point. By the definition of $\lambda(\cdot)$, we see that these first $\lambda(IV)$ blocks bring us to a cyclic node in the functional graph $G_{f_B}$. After the first $\lambda(IV)$ blocks, message $M_0$ ends immediately, while $M_1$ has $\mu(IV)$ more blocks to complete a full additional cycle in the functional graph. Starting from $IV$ in $G_{f_B}$, traversing $\lambda(IV) + \mu(IV)$ out-edges brings us to the same node as if we only traverse $\lambda(IV)$ out-edges. Therefore, when $M_0$ and $M_1$ are passed to our unspecified hash function, the two will collide directly prior to the final padding block.

The above collision construction still works if we replace $\lambda(IV)$ with any $a \geq \lambda(IV)$, as any such value is sufficient to bring $M_0$ and $M_1$ into a cycle in the functional graph. Furthermore, after $M_0$ and $M_1$ arrive at the same cyclic node, the two can traverse the cycle any whole number of times and still collide. Thus, we see that the messages
	\[M_0 = [B]^{a + c_1 \mu(IV)}, ~~~~~ M_1 = [B]^{a + c_2 \mu(IV)}\]
also collide for any $a \geq \lambda(IV)$ and any natural numbers $c_1, c_2$. In fact, for fixed $a \geq \lambda(IV)$, the set
	\[\left\{[B]^{a + c \mu(IV)} ~ \left| ~ c \in \N \right.\right\}\]
is an infinite family of messages which mutually collide up until the final padding blocks are processed.

There are two reasons that we do not yet have a full collision formula. First, it is nontrivial to compute $\lambda(IV)$ and $\mu(IV)$. From the results cited in the previous section, these values are roughly $2^{\ell/2}$ each, and we tend to have $\ell \geq 128$ in practice. Second, the mutually colliding messages all have different lengths; hence, they are padded differently, so we should expect the collisions to disappear once their final blocks are processed.

We begin by addressing the problem of computing $\lambda(IV)$ and $\mu(IV)$. While we should expect it to be infeasible in general to compute $\lambda(IV)$, we do not need an exact value---it is sufficient simply to find a value $a \geq \lambda(IV)$. Since our functional graph $G_{f_B}$ only has $2^\ell$ nodes, we know that $\lambda(IV) < 2^\ell$, hence the choice $a = 2^\ell$ will suffice.

Similarly, we do not need to know the exact value of $\mu(IV)$---we will still have an infinite family of mutually colliding messages if we replace the term $c \mu(IV)$ with any multiple of $\mu(IV)$. And while computing $\mu(IV)$ is probably infeasible, we know that $\mu(IV) \in \{1, 2, \dots, 2^\ell\}$. Thus, $(2^\ell)!$ must be a multiple of $\mu(IV)$. We therefore obtain our formula that produces hash collisions:
	\[[B]^{2^\ell + c(2^\ell)!} \tag{*}\]
For any fixed block $B$, compression function $f_B$, and any two values of $c \in \N$, we obtain a hash collision. Indeed, the set
	\[\left\{[B]^{2^\ell + c(2^\ell)!} ~|~ c \in \N\right\}\]
is a subset of the infinite family of mutually colliding messages. As previously mentioned, some padding schemes may thwart this formula. We have actually already resolved this issue for some functions, however. The specification for MD5, given in RFC 1321 \cite{md5}, explicitly allows messages of arbitrary length: if the input length is at least $2^{64}$ bits, then the padding only uses the low order 64 bits of the length in the padding. Therefore, for two messages to receive the same MD5 padding, it is not necessary that they have the same length---it is sufficient that their lengths be equivalent modulo $2^{64}$. For MD5, all messages produced by our formula have bit lengths that are multiples of $2^\ell = 2^{128}$, so their lengths are all equivalent modulo $2^{64}$. Hence, our formula produces full collisions for MD5.

SHA1 and SHA256 have, respectively, $\ell = 160$ and $\ell = 256$. This means that for these functions, all messages produced by our formula have equivalent lengths modulo $2^{64}$. However, the NIST standardizations of these functions, given in FIPS 180-4 \cite{fips180}, require input messages to have fewer than $2^{64}$ bits. (This is not unreasonable---it is still difficult to envision a future in which files with $2^{64}$ bits are hashed. And allowing such long inputs allows for some potent attacks \cite{kelsey}.) All messages produced by our formula for SHA1 and SHA256 have lengths much greater than $2^{64}$. For these functions, the formula only produces collisions on the iterated compression function. If the domains of SHA1 and SHA256 were extended in the natural way, our formula would produce full collisions for these functions, too.

\section{Analysis of Message Lengths}

The messages produced by our formula are too long for SHA1 and SHA256, but how long are they, really? The dominant term in the formula is $(2^\ell)!$, which is not a familiar quantity. We can, however, set upper and lower bounds on it. The upper bound is obtained by replacing all $2^\ell$ individual factors with $2^\ell$, while the lower bound is obtained by neglecting the smallest half of the terms and replacing all remaining terms with $2^\ell/2 = 2^{\ell-1}$. We obtain
	\[(2^{\ell-1})^{2^{\ell-1}} < (2^\ell)! < (2^\ell)^{2^\ell}.\]
Rewriting these bounds as
	\[2^{2^{\ell-1+\log(\ell-1)}} < (2^\ell)! < 2^{2^{\ell+\log \ell}},\]
it becomes clear that $(2^\ell)!$ is double exponential in $\ell$. The number of $b$-bit blocks in our messages is roughly $c(2^\ell)!$ (the term $2^\ell$ is comparatively negligible), so the total bit length of our messages is roughly
	\[bc2^{2^\ell}\]
for arbitrary choice of $c \in \N$. For typical values of $\ell$ with $\ell \geq 128$, it is safe to say that such a message cannot ever be fully hashed, much less written out in full. (The possible exception is the message with $c = 0$, but that message has length $b2^\ell$, which is still prohibitively large for the foreseeable future.)

Our collision formula is perhaps excessive---we upper bounded the tail and cycle lengths in functional graphs by $2^\ell$, but by theorem 2, the maximum tail and cycle lengths over the whole graph are on the order of $2^{\ell/2}$ in expectation. This suggests a collision formula that still works in expectation but produces smaller messages:
	\[[B]^{\lceil 0.78 \cdot 2^{\ell/2} \rceil + c(\lceil 1.74 \cdot 2^{\ell/2} \rceil)!}\]
While the messages produced by this formula are objectively much shorter than those produced by the first formula, these message lengths are still double exponential in $\ell/2$. This does not bring us meaningfully closer to practical message lengths. And if we further decrease the message lengths, we correspondingly decrease the probability that the formula yields collisions.

Our formula exploits cycles in functional graphs to find collisions. This method, at minimum, requires us to enter a cycle. By theorem 1, we know that the average node is at distance $\sqrt{\pi 2^\ell/8} \approx 0.63 \cdot 2^{\ell/2}$ from the nearest cycle, so it is difficult to see how this method could produce colliding messages much shorter than about $2^{\ell/2}$ bits.

The cycle lengths pose an even greater problem. Whereas our formula only needs an upper bound on $\lambda(IV)$, it needs an exact multiple of $\mu(IV)$, and it is not obvious that we can find a small multiple of $\mu(IV)$ without determining $\mu(IV)$. Since $\mu(IV)$, like $\lambda(IV)$, has an expected value of about $0.63 \cdot 2^{\ell/2}$, determining $\mu(IV)$ in the straightforward way puts us roughly even with a randomized birthday collision attack---and the colliding messages found this way would still be too long for SHA1 and SHA256.

One idea for a cycle-based collision attack (and perhaps also a second preimage attack) is to hope that, for our block choice $B$, the node $IV$ is a cyclic node. However, there are only about $\sqrt{\pi 2^\ell/2}$ cyclic nodes out of $2^\ell$ total nodes, and it is unclear how to efficiently identify cyclic nodes.

Our formula identifies long colliding messages with virtually no work. It does not seem, however, that there is a noteworthy length/time tradeoff here. Collisions found with this methodology necessarily have at least exponential length in expectation, and achieving single exponential length straightforwardly requires exponential time.

\section{Collision Resistance}

While the messages produced by our collision formula are unimaginably long, the very existence of a hash collision formula raises an interesting question: does this formula technically break the collision resistance of some hash functions?

Existing definitions of collision resistance are just vague enough that the answer is arguably yes. Collision resistance is formalized in terms of keyed hash functions and a ``collision game''. Keyed hash functions differ from real world hash functions in that they cannot be evaluated without knowledge of a special key. Real world hash functions can be viewed as keyed hash functions with standardized key values.

In a collision game \cite{gb} \cite{katz}, a key is randomly generated, and an adversary must produce messages which collide for that particular key. Roughly speaking, a hash function is said to be collision resistant if its keyed variant thwarts the adversary with overwhelming probability. There are several versions of the collision experiment---while the adversary is usually given the key immediately, several variants \cite{gb} require the adversary to produce one or both of the colliding messages before being given the key.

The adversary in the collision game must run in time polynomial in the security parameter. A message of double exponential length requires double exponential time to write out in full. However, the long colliding messages produced by our collision formula are highly redundant, consisting of the same block repeated many times. These long colliding messages can be compressed to logarithmic space, $O(\log \ell + \log c)$, by simply writing out formula (*) for two different values of $c$. The collision resistance definitions given by \cite{gb} and \cite{katz} do not state that the colliding messages must be of polynomial length, and they only state that the adversary must ``output'' the colliding messages. Since these definitions do not explicitly state that the adversary must write out the messages in full, it is arguably acceptable for the adversary to provide compressed output.

If compressed output is allowed, then the adversary can win the collision game with probability 1 under all models, including those which require the adversary to produce collisions without knowledge of the chosen key. This is because our formula is very general---it does not use any information about hash function internals or key values. It only takes advantage of inevitable properties of functions over finite domains.

\section{Conclusion}

This hash collision formula is a theoretical curiosity. It does not compromise collision resistance in any practical sense, and there is good reason to believe that this methodology cannot be converted into a practical collision attack.

Nevertheless, it is theoretically ambiguous whether the formula compromises the security of real world hash functions. If the adversary in the collision game is allowed to give compressed output, then the formula trivially breaks the collision resistance of MD5 and nearly does the same to SHA1 and SHA256. A collision attack on MD5 is perhaps unexciting, since collisions can already be found in about $2^{24.1}$ evaluations of the compression function \cite{md5coll}. Our collision formula, however, produces collisions without any evaluations of the compression function.

This collision attack is impractical, and can be theoretically ruled out by tightening the definition of collision resistance. SHA1 and SHA256 already accomplish this by placing limits on input message length. This defense can be extended to all hash functions by specifying that adversaries must write out colliding messages in full, or by requiring that the colliding messages must have length polynomial in the security parameter.

\section*{Acknowledgement}

This work was supported by the National Science Foundation (grant CCF-1420750). The author is also grateful to Eric Bach for many helpful discussions over the course of this research.

\end{document}